\documentclass[usenatbib]{mn2e}

\newcommand{\msun}{M$_{\odot}$\,}
\newcommand{\apj}{ApJ}
\newcommand{\apjs}{ApJS}
\newcommand{\apjl}{ApJ}
\newcommand{\aap}{A\&A}
\newcommand{\mnras}{MNRAS}
\newcommand{\araa}{ARA\&A}
\newcommand{\pasp}{PASP}
\newcommand{\rmxaa}{Rev. Mexicana Astron. Astrofis.}
\newcommand{\procspie}{Proc.\ SPIE}
\newcommand{\aj}{AJ}
\newcommand{\nodata}{...}
\usepackage{epstopdf}
\usepackage{graphicx}
\usepackage{amsmath}
\usepackage{longtable}

\title{Stellar Multiplicity and Debris Disks: An Unbiased Sample}

\author[D. R. Rodriguez et al.]{
David R. Rodriguez$^{1}$\thanks{E-mail: drodrigu@das.uchile.cl}, 
Gaspard Duch{\^e}ne$^{2,3}$, Henry Tom$^{2,4}$, Grant Kennedy$^{5}$, \newauthor
Brenda Matthews$^{6,7}$, Jane Greaves$^{8}$, Harold Butner$^{9}$ \\
$^{1}$Departamento de Astronom{\'i}a, Universidad de Chile, Casilla 36-D, Santiago, Chile\\
$^{2}$Department of Astronomy, UC Berkeley, 501 Campbell Hall, Berkeley CA 94720-3411, USA\\
$^{3}$Univ. Grenoble Alpes/CNRS, IPAG, F-38000 Grenoble, France\\
$^{4}$Univ. California San Francisco / SD, PRDS, 707 Parnassus Avenue, San Francisco, CA 94143-0758 USA\\
$^{5}$Institute of Astronomy, University of Cambridge, Madingley Road, Cambridge CB3 0HA, UK\\
$^{6}$National Research Council of Canada Herzberg Astronomy \& Astrophsyics, 5071 W. Saanich Road, Victoria, BC, V9E 2E7, Canada\\
$^{7}$Department of Physics \& Astronomy, University of Victoria, 3800 Finnerty Rd, Victoria, BC, V8P 5C2, Canada\\
$^{8}$SUPA, School of Physics and Astronomy, University of St.\ Andrews, North Haugh, St. Andrews KY16 9SS, UK\\
$^{9}$Department of Physics \& Astronomy, James Madison University, Harrisonburg, VA, USA
}
\begin{document}


\pagerange{\pageref{firstpage}--\pageref{lastpage}} \pubyear{2015}

\maketitle

\label{firstpage}


\begin{abstract}
Circumstellar dust disks have been observed around many nearby stars. However, many stars are part of binary or multiple stellar systems. A natural question arises regarding the presence and properties of such disks in systems with more than one star. 
To address this, we consider a sample of 449 systems (spectral types A--M) observed with the {\it Herschel Space Observatory} as part of the DEBRIS program. We have examined the stellar multiplicity of this sample by gathering information from the literature and performing an adaptive optics imaging survey at Lick Observatory. Five new companions were revealed with our program. In total, we identify 188 (42\%) binary or multiple star systems. The multiplicity of the sample is examined with regards to the detection of circumstellar disks for stars of spectral types AFGK.

In general, disks are less commonly detected around binaries than single stars, though the disk frequency is comparable among A stars regardless of multiplicity. However, this sample reveals the period distribution of disk-bearing binaries is consistent with that of non-disk binaries and with comparison field samples. We find that the properties of disks in binary systems are not statistically different from those around single stars.
Although the frequency of disk-bearing FGK binaries may be lower than in single star systems, the processes behind disk formation and the characteristics of these disks are comparable among both populations.
\end{abstract}

\begin{keywords}
binaries: general --- infrared: stars --- planetary systems: formation.
\end{keywords}

\section{Introduction}

Disks rich in gas and dust around young stars are sites of planet formation.
Jovian-class planets need to form rapidly, as gas within the disks dissipates over a period of a few million years \citep{Zuckerman:1995,Pascucci:2006}.
The dust in the system is also removed eventually, either through accretion onto larger objects, stellar winds, or radiative processes \citep{Haisch:2001,Uzpen:2009}. However, collisions between any planetesimals in the disk can generate a second population of dust. These second-generation systems are known as debris disks and generally contain very little gas \citep[and references therein]{Zuckerman:2001,Wyatt:2008}.
Our own solar system's Kuiper belt may be analogous to these debris disks \citep[e.g.,][and references therein]{Luu:2002}.
The study of disks, from the gas-rich protoplanetary disks to the gas-poor debris disks, is necessary for a more complete understanding of the formation and evolution of planetary systems.

Approximately half of all stars are in binary or multiple star systems \citep{Duquennoy:1991,Eggleton:2008,Raghavan:2010,Duchene:2013}.
Given how common binary stars are, it is important to address the properties of planet formation in such common systems.
About 20\% of known extra solar planets reside in wide binaries with separations of order 100s of AU \citep{Raghavan:2006,Eggenberger:2007}.
The {\it Kepler} satellite has also revealed several planets orbiting pairs of close binary systems \citep[e.g.,][]{Doyle:2011,Welsh:2012,Orosz:2012a, Orosz:2012b, Schwamb:2013,Kostov:2014}. 
Earlier work with eclipse timing variations have suggested planets around the binaries HW~Virginis, CM~Draconis, and NN~Serpentis (a post-common envelope binary) \citep{Deeg:2008,Lee:2009,Beuermann:2010}.

Circumstellar disks can be used as indirect evidence of planets in binary systems.
For example, the close (3.4-day) main sequence binary BD~+20~307 displays a large amount of warm dust in the terrestrial planet zone \citep{Song:2005}, which can be interpreted as the result of a planetary-scale collision in this $\sim$1~Gyr old binary system \citep{Zuckerman:2008b, Weinberger:2011}.
Disk studies among young pre-main sequence systems have shown that the presence of a nearby companion star can readily truncate and disperse disks \citep{Jensen:1996,Bouwman:2006,Cieza:2009,Andrews:2010,Kraus:2012}, in good agreement with what is expected from numerical simulations \citep{Lubow:2000,Artymowicz:1994}. 
Older debris disk binary systems also show a similar behavior \citep{Trilling:2007,Rodriguez:2012}.
Studying circumstellar and circumbinary disks is then a complementary way to explore the properties of planet formation in binary star systems.

The {\it Herschel Space Observatory}\footnote{{\it Herschel} is an ESA space observatory
with science instruments provided by European-led Principal Investigator consortia and with important participation from NASA.} \citep{Pilbratt:2010} offers the best opportunity to date of performing a large scale volume-limited survey for debris disks, the ideal approach to systematically analyze the interplay between debris disks and stellar multiplicity.
The DEBRIS (Disc Emission via a Bias-free Reconnaissance in the Infrared/Submillimetre; \citealt{Matthews:2010}) program has explored nearby star systems to search for infrared excesses indicative of disks at wavelengths longward of 70\micron.
Prior work on the multiplicity of debris disk stars has been limited by small samples with strong biases; for example, considering only debris disk systems \citep{Rodriguez:2012} or only binary and multiple star systems \citep{Trilling:2007}. 
The DEBRIS sample has observed stars regardless of prior known disks or stellar multiplicity and thus provides a better sample in which to explore the relationships between these two phenomena.
Several DEBRIS binaries/multiples with disks have been individually explored in detail \citep{Kennedy:2012a,Kennedy:2012b,Kennedy:2014}.
In this paper, we explore the multiplicity statistics of the DEBRIS sample and its role in the detection of circumstellar and circumbinary disks.

\section{Multiplicity in the DEBRIS Sample} \label{debris:sample}

\begin{table*}
\centering
\begin{minipage}{120mm}
\caption{DEBRIS Sample}\label{tab:data1}
\begin{tabular}{llcp{6cm}p{3cm}}
\hline
Name & UNS ID$^a$ & Multiple? & Notes & References \\
\hline
HD 38	&	K050	&	Y	&	(8.2K6+9.9M2; 6.041'')	&	HIP, SHA11	\\
HD 166	&	G030	&	N	&	optical double; LAF07: background source 10.23" away	&	WDS, LAF07, KIY08	\\
HD 693	&	F069	&	N	&	F7V	&	ET08, TAN10	\\
HD 739	&	F096	&	N	&	F4V                                                                                                                        	&	ET08	\\
HD 1237	&	G070	&	Y	&	(G6+M4; 3.857")	&	CHA06	\\
HD 1326	&	M011	&	Y	&	(8.31M2V+11.36M6V; 2600y, 41.15") WDS: AC pair ruled out; TAN10: companions to A \& B found, but no common motion	&	WDS, VB6, TAN10	\\
HD 1404	&	A103	&	Y	&	A2V, proper motion companion found in Lick AO images	&	ET08, ThisWork	\\
HD 1581	&	F005	&	N	&	F9V                                                                                                                        	&	ET08	\\
HD 1835	&	G118	&	N	&	optical double, different proper motions	&	WDS, RAG10	\\
HD 2262	&	A021	&	N	&	A7V                                                                                                                        	&	ET08	\\
HD 3196	&	F092	&	Y	&	(5.61(F8V + ?; 2.082d) + 6.90G0V; 6.89yr e=0.76)                                                                            	&	ET08	\\
HD 3443	&	G044	&	Y	&	(6.37G8V + 6.57G9V; 25.09yr e=0.24)                                                                                         	&	ET08	\\
HD 3651	&	K045	&	Y	&	(4.55+16.87T7.5; 42.9")  WDS: AB optical, AC common proper motion	&	WDS, KIY08, LUH07	\\
HD 4391	&	G041	&	Y	&	(5.8G1V + (13.5; 16.6Ó 251AU) + (14M5; 49Ó 742AU)); ET08 incorrectly lists as single, WDS entries somewhat contradictory	&	WDS, RAG10	\\
HD 4628	&	K016	&	N	&	K2V                      	&	ET08	\\
HD 4676	&	F124	&	Y	&	((F8V + F8V; 13.82d  e=0.24) + ?; 0.25")       	&	ET08	\\
HD 4747	&	G089	&	Y	&	(?+?; 6832d e=0.64)	&	SB9	\\
HD 4813	&	F038	&	N	&	F7IV-V	&	ET08, TAN10	\\
HD 4967	&	K127	&	Y	&	(8.16K5+15.2M5.5; 16.8")	&	WDS, POV94, HAW97, GOU04	\\
HD 5133	&	K089	&	N	&	no stellar companion	&	RAG10	\\
HD 5448	&	A090	&	N	&	A5V                   	&	ET08	\\
\hline
\end{tabular}\\
{Systems in the DEBRIS sample. The notation follows that of \citet{Eggleton:2008}.
The full table is available online. Details on binary and multiple systems are listed in Table~\ref{tab:data2}. Systems with multiplicity flag ``A'' have some indication that they are astrometric binaries, yet no other information is available. \\
$^a$: Unbiased Nearby Stars from \citet{Phillips:2010}\\
Table references as follows: 
HIP	:	Hipparcos Catalog (\citealt{HIP}),
SB9	:	The ninth catalogue of spectroscopic binary orbits (\citealt{Pourbaix:2004}),
VB6	:	Sixth Catalog of Orbits of Visual Binary Stars (\citealt{Hartkopf:2001}),
WDS	:	Washington Double Star Catalog (\citealt{Mason:2001}),
BAI10	:	\citet{Baines:2010},
BAR09	:	\citet{Bartlett:2009},
BER10	:	\citet{Bergfors:2010},
CAB07	:	\citet{Caballero:2007},
CAR05	:	\citet{Carson:2005},
CAR11	:	\citet{Carson:2011},
CHA06	:	\citet{Chauvin:2006},
CLO07	:	\citet{Close:2007},
DEL99	:	\citet{Delfosse:1999},
DER11	:	\citet{deRosa:2011},
DER12	:	\citet{deRosa:2012},
DM91	:	\citet{Duquennoy:1991},
EGG07	:	\citet{Eggenberger:2007},
EHR10	:	\citet{Ehrenreich:2010},
EIS07	:	\citet{Eisenbeiss:2007},
EKE08	:	\citet{Eker:2008},
ET08	:	\citet{Eggleton:2008},
FRA07	:	\citet{Frankowski:2007},
FUH14	:	\citet{Fuhrmann:2014},
GOL06	:	\citet{Goldin:2006},
GOL07	:	\citet{Goldin:2007},
GOU04	:	\citet{Gould:2004},
HAR12	:	\citet{Hartkopf:2012},
HAW97	:	\citet{Hawley:1997},
HIN02	:	\citet{Hinz:2002},
HIN10	:	\citet{Hinkley:2010},
JAO03	:	\citet{Jao:2003},
KIY08	:	\citet{Kiyaeva:2008},
KOH12	:	\citet{Kohler:2012},
KON10	:	\citet{Konopacky:2010},
LAF07	:	\citet{Lafreniere:2007},
LAG06	:	\citet{Lagrange:2006},
LAW08	:	\citet{Law:2008},
LEC10	:	\citet{Leconte:2010},
LEI01	:	\citet{Leinert:2001},
LEI97	:	\citet{Leinert:1997},
LEP07	:	\citet{Lepine:2007},
LUH07	:	\citet{Luhman:2007},
MAK05	:	\citet{Makarov:2005},
MAM10	:	\citet{Mamajek:2010},
MAM13	:	\citet{Mamajek:2013},
MAR14	:	\citet{Marion:2014},
MAS98	:	\citet{Mason:1998},
MCA93	:	\citet{McAlister:1993},
MCC04	:	\citet{McCarthy:2004},
MET09	:	\citet{Metchev:2009},
NAK94	:	\citet{Nakajima:1994},
NAK95	:	\citet{Nakajima:1995},
NID02	:	\citet{Nidever:2002},
NIE10	:	\citet{Nielsen:2010},
OPP01	:	\citet{Oppenheimer:2001},
PHI10	:	\citet{Phillips:2010},
POT02	:	\citet{Potter:2002},
POV94	:	\citet{Poveda:1994},
RAG10	:	\citet{Raghavan:2010},
ROB11	:	\citet{Roberts:2011},
SCH07	:	\citet{Schroder:2007},
SCH11	:	\citet{Schneider:2011},
SEG00	:	\citet{Segransan:2000},
SHA11	:	\citet{Shaya:2011},
TAN10	:	\citet{Tanner:2010},
TOK08	:	\citet{Tokovinin:2008},
TOK10	:	\citet{Tokovinin:2010},
TOK11	:	\citet{Tokovinin:2011},
TOK92	:	\citet{Tokovinin:1992},
WIL01	:	\citet{Wilson:2001}.
} 
\end{minipage}
\end{table*}

The DEBRIS sample consists of 451 stars, of which 2 are not primaries (Fomalhaut B \&C; see \citealt{Mamajek:2013,Kennedy:2014}), and includes the 5 systems observed by the Herschel Guaranteed Time (GT) disks program (PI: Olofsson). The majority of the stars were observed by Herschel as part of the DEBRIS program, with some observations coming from the DUNES program \citep{Eiroa:2010}. 
The sample of 449 primaries is roughly equally divided in spectral types A  through M and is volume-limited for each individual spectral type, where the limit is 
45.5, 23.6, 21.3, 15.6, and 8.6~pc 
for A, F, G, K, and M spectral types, respectively \citep{Phillips:2010}.
Herschel observations with PACS \citep{Poglitsch:2010} and SPIRE \citep{Griffin:2010} have been performed for the DEBRIS sample in order to search for and characterize far-IR emission from circumstellar (or circumbinary) dust.
Except for cases in which a companion fell within the field of view, we didn't explicitly aim to observe companions with exception of Fomalhaut's companions, of which companionship was unknown until recently (see \citealt{Mamajek:2013,Kennedy:2014}). 
Thus, we may be missing some circumsecondary disks and thus focus on circumprimary or circumbinary disks only.
Simultaneously to these Herschel observations, we have gathered literature data and adaptive optics (AO) observations (see Section~\ref{AOSearch}) in order to characterize the multiplicity of stars in this sample.

\begin{figure}
\begin{center}
\includegraphics[width=88mm,angle=0]{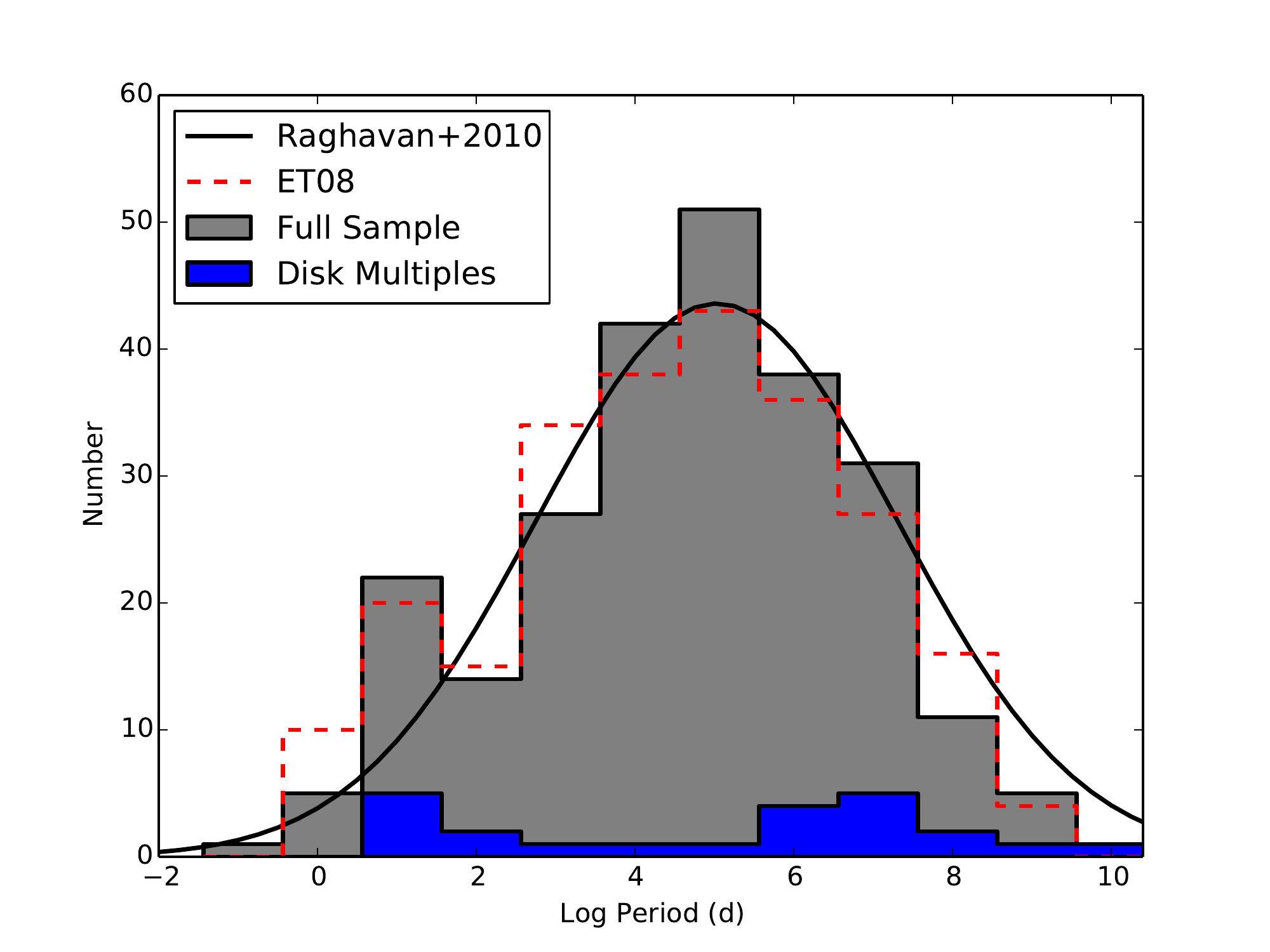}
\end{center}
\caption{Distribution of calculated and measured orbital periods for multiples (gray histogram). Triples and higher order multiples have more than one period per system. The dashed line is the period distribution from \citet{Eggleton:2008} and the solid curve is from \citet{Raghavan:2010}. 
Both are normalized to the number of periods in the DEBRIS sample. The blue shaded histogram shows the distribution of periods for disk-bearing multiples in our sample. }
\label{fig:period}
\end{figure}

\begin{table*}
 \centering
 \begin{minipage}{120mm}
  \caption{DEBRIS Multiples}
  \begin{tabular}{lccccccc}
\hline
Name & Component & Period (d) & Sep (AU) & Eccentricity & Derived$^a$ & Dust R (AU) \\
\hline
HD 38	&	AB	&	1.95E+05	&	68.2	&		&	P	&		\\
HD 1237	&	AB	&	1.91E+05	&	67.5	&		&	P	&		\\
HD 1326	&	AB	&	9.50E+05	&	147.3	&		&	N	&		\\
HD 1404	&	AB	&	1.10E+06	&	272	&		&	P	&	26.4	\\
HD 3196	&	AB	&	2.082	&	0.035	&		&	a	&		\\
	&	AB-C	&	2.52E+03	&	3.95	&	0.76	&	a	&		\\
HD 3443	&	AB	&	9.16E+03	&	8.93	&	0.24	&	a	&		\\
HD 3651	&	AB	&	3.72E+06	&	474	&		&	P	&		\\
HD 4391	&	AB	&	1.23E+06	&	252	&		&	P	&		\\
	&	AC	&	6.26E+06	&	743	&		&	P	&		\\
HD 4676	&	AB	&	13.82	&	0.126	&	0.24	&	a	&		\\
	&	AB-C	&	4.40E+03	&	5.86	&		&	P	&		\\
HD 4747	&	AB	&	6.83E+03	&	7.23	&	0.64	&	a	&		\\
HD 4967	&	AB	&	1.72E+06	&	262	&		&	P	&		\\
HD 7439	&	AB	&	9.13E+06	&	1162	&		&	P	&		\\
 \hline
\end{tabular}\\
{Measurements of components of binaries and multiples in the DEBRIS sample. The location of blackbody dust grains for dusty systems is also listed. For HD~216956, dust R is listed first for A, then C. For HD~223352, the AC pair's dust is located around the C component. The full table is available online. 
For clarity, we have rounded periods to 1 day when larger than 100 days. Similarly, separations in AU are rounded to 0.1 AU when larger than 20AU and to 1 AU when larger than 200 AU. We refer readers to the references listed in Table~\ref{tab:data1} for the most exact values. \\
$^a$: Quantity derived: P, period; a, semimajor axis; N, none. See Section~\ref{debris:sample} for details. } \label{tab:data2}
\end{minipage}
\end{table*}

\begin{table*}
 \centering
 \begin{minipage}{120mm}
  \caption{Multiplicity Fractions\label{debris:binfraction}}
  \begin{tabular}{cccccc}
  \hline
{Sp.\ Type} & {Number} & {This Work} & {From RZ12}
& {From ET08} & Others\\
 \hline
B	& \nodata & \nodata & $83^{+6}_{-23}$ & \nodata & \nodata \\
A	& $35/86$ & $41\pm{5}$ & $26^{+7}_{-5}$ & 46.0 & $>$50 \\ 
F	& $50/94$ & $53\pm{5}$ & $12^{+7}_{-3}$ & 47.4 & \nodata \\ 
G	& $43/90$ & $48\pm{5}$ & $33^{+15}_{-10}$ & 45.0 & 44 \\
K	& $37/91$ & $41\pm{5}$ & $40^{+21}_{-16}$ & 29.1 & \nodata \\ 
M	& $23/88$ & $26^{+5}_{-4}$ & \nodata & \nodata & 26 \\
All	& $188/449$ & $42\pm4$     & $25\pm4$ &  42.8 & \\ 
\hline
\end{tabular}\\
{Fraction of multiple stars (all as percentages) broken down by spectral type of the primary star with comparisons from the literature (RZ12: \citealt{Rodriguez:2012}; ET08: \citealt{Eggleton:2008}; \citealt{Raghavan:2010}, \citealt{Duchene:2013}). Note that the \citet{Rodriguez:2012} sample only contains disk-bearing systems. 
Errors are binomial errors estimated as described by \cite{Burgasser:2003}.
}\label{tab:bin1}
\end{minipage}
\end{table*}

Given that the majority of the stars in the DEBRIS sample are close to Earth and well-studied in the literature, published data concerning stellar multiplicity exists for a large fraction of the sample. Our AO observations complement this literature search (see Section~\ref{AOSearch}).
We present the full list of stars and multiplicity information in Table~\ref{tab:data1}.
Some systems have tentative evidence of being astrometric binaries, but no additional information exists to confirm this. These systems are treated as single stars for this study.
Binaries and higher multiples are listed with more details in Table~\ref{tab:data2}. 
In general, we only have measurements of the period of spectroscopic binaries and projected separation for visual binaries. We derive the remaining quantities (ie, period or semimajor axis) assuming the measured separation is the orbital semimajor axis, orbits are circular, and adopting main-sequence masses from \citet{Baraffe:1998} and \citet{Siess:2000}. Which quantity is derived is indicated in Table~\ref{tab:data2}.
While clearly not all orbits are circular and projection effects have not been incorporated in our analysis (corrections for these are small, of order $\sim$10\%; see \citealt{Dupuy:2011}), this nevertheless provides useful information in a statistical manner.  

The sample has 188 (42\%) star systems where two or more stars are present. 
As previously mentioned, we calculate or adopt literature values for parameters of the system, such as the orbital period and semi-major axis.
In Figure~\ref{fig:period}, we show the period distribution for all multiples in the sample, including those determined from our AO observations (Section~\ref{AOSearch}). This includes all periods, so, for example, triples have two periods counted. 
Also shown is the period distributions from \citet{Raghavan:2010} and \citet{Eggleton:2008} normalized to the same number of periods. These two latter distributions sample solar-type stars and bright systems and are representative of what we would expect for our sample.
The distributions are remarkably similar to each other suggesting that our multiplicity survey is not biased against any particular range of periods.

The multiplicity fraction among A--K stars in our sample is 40--50\%, consistent with some prior studies \citep{Eggleton:2008, Duquennoy:1991}. 
However, \citet{Duchene:2013} estimate a lower limit to the multiple star fraction among A stars of 50\% suggesting we may still be missing some A-star binaries.
Among M-dwarfs we find a multiplicity fraction of 26\%, in agreement with the multiplicity fraction listed in \citet{Duchene:2013}. 
However, we note that these faint stars have not been as intensely studied for binarity as more massive stars in the DEBRIS sample. 
Furthermore, the DEBRIS Herschel sensitivity towards disks around M-dwarfs is also low (see Figure~2 in \citealt{Matthews:2014}). 
The combination of low detection rates (multiplicity or debris disks) results in small number statistics whose robustness is too limited for a detailed analysis. 
Therefore, we primarily focus on the A--K sample in this article.
We summarize the multiplicity statistics, divided by spectral type, in Table~\ref{tab:bin1}.

\section{Adaptive Optics Search for Companions}\label{AOSearch}

\subsection{Survey Setup and Observations}

We have utilized the adaptive optics (AO) camera IRCAL \citep{Lloyd:2000,Perrin:2008} at Lick Observatory to search for companions separated from the primary by $\sim$10--1000~AU.
The NIR camera offers a $\approx$0\farcs077 pixel scale (which provides Nyquist sampling for diffraction-limited observations at 2.2 micron on the Shane 3m telescope) and a 20\arcsec\ FOV. Precise pixel scale (including the known slight anamorphism of the camera) and orientation was determined by observations of multiple known binaries. The precision is estimated to be 1\% for the pixel scale and 0.7deg for the absolute orientation.
The dither pattern typically used was a 5\arcsec-on-the-side square, giving us full coverage out to 12\farcs5 from the primary, and we only consider companions within this radius in the survey.
Tables~\ref{binarytab1} and \ref{singletab1} list our observations and measured parameters for binary and single stars, respectively.

\begin{figure}
\begin{center}
\includegraphics[width=88mm,angle=0]{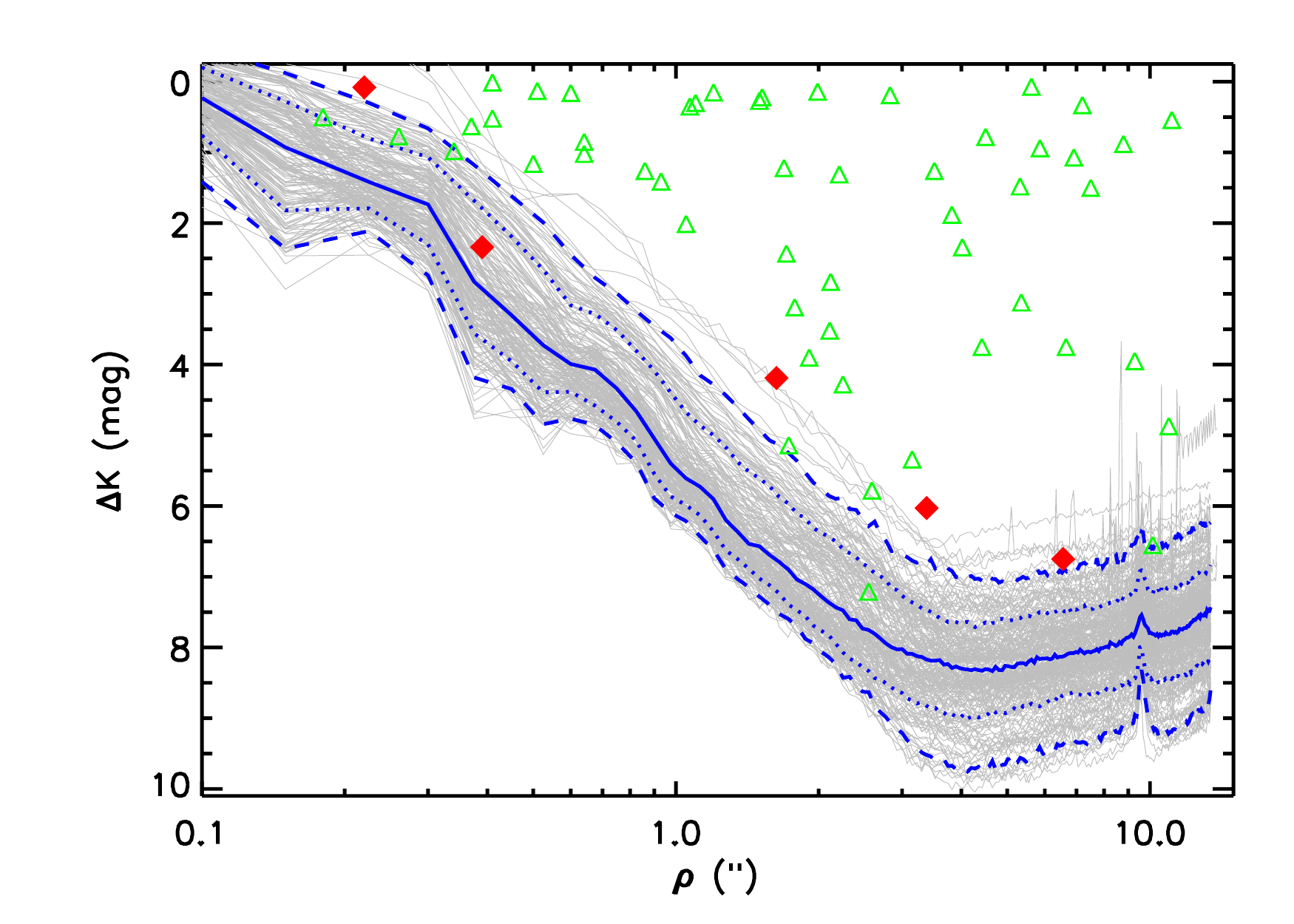}
\end{center}
\caption{IRCAL 5-$\sigma$ detection limits for the sample including detected known systems (green triangles) and new companions (red diamonds). In blue we show the median 5-$\sigma$ sensitivity of our survey, along with the $\pm$1 and $\pm$2-$\sigma$ ranges.}
\label{fig:licklim}
\end{figure}

We carried out our observations on various nights in 2009 June \& October, 2010 August, and 2012 March \& September.
A total of 221 DEBRIS targets were observed over this time period. This corresponds to 75\% of all targets with declinations larger than --10 degrees.
For our observations, we used a dithered sequence to remove artifacts and cosmic rays and observed with either the Ks or Br-$\gamma$ filters. Objects that had a close companion or appeared extended were subsequently observed at J and either H or FeII in order to estimate colors and spectral types. The choice of FeII and Br-$\gamma$ over H and Ks was due to saturation on bright targets. The FeII and Br-$\gamma$ filters have central wavelengths of 1.644 and 2.167\AA\ with bandwidth FWHM of 0.016 and 0.020\AA, respectively. We assume the flux ratio between the primary and any companion at Br-$\gamma$ is comparable to that in Ks (and similarly for FeII and H).
To confirm companions we either obtained a second epoch and tested for common motion, or estimated the spectral type of the companion with our color information and verified that the photometric distance agrees with that of the primary.
We achieve a typical contrast of 6 magnitudes (5-$\sigma$) at separations $>$1 arcseconds (see Figure~\ref{fig:licklim}). 
The \citet{Siess:2000} models, for ages up to 1~Gyr, predict a flux ratio of 9--9.5~mag between A0 stars (2.5--3\msun, depending on age) and 0.1\msun stars at K band. Our median detection limit beyond $\sim$2\arcsec\ is 8 magnitudes suggesting we are close to being complete for stellar mass companions.
While a few lower mass companions may remain to be found, particularly around the later spectral types, we do not anticipate that these would significantly change the results. Our AO results are incorporated in Figure~\ref{fig:period} in which we see no strong bias or incompleteness against any particular range of orbital periods.

\begin{table}
 \centering
\caption{IRCAL Companion Measurements}
\begin{tabular}{llrrrlc}
\hline
Name & Filter & Sep  & PA 	& $\Delta$m 	& UT Date & Depth \\
	  &	      & ($''$) & (deg)	& (mag)		&		 & (mag) \\
\hline
\hline
HD 38	&	Ks	& 5.62 & 5.87 & 0.071 & 	2009-10-28	&	8.47	\\
HD 1404	&	Ks	& 6.60 & 144.40 & 6.682 & 	2009-10-28	&	8.59	\\
		&	BrG	& 6.56 & 144.23 & 6.746 & 	2012-09-28	&		\\
HD 3196	&	Ks	& 0.26 & 248.43 & 0.770 & 	2010-08-03	&	7.97	\\
HD 16160	&	BrG	& 1.73 & 318.49 & 5.142 & 	2010-08-05	&	9.13	\\
HD 16765	&	Ks	& 3.82 & 300.81 & 1.880 & 	2010-08-04	&	9.06	\\
HD 16970	&	Ks	& 2.21 & 297.83 & 1.312 & 	2010-08-04	&	8.93	\\
HD 19994	&	BrG	& 2.12 & 203.15 & 2.830 & 	2009-10-28	&	8.36	\\
HD 56537	&	BrG	& 9.29 & 35.72 & 3.953 & 	2009-10-28	&	8.42	\\
HD 56986	&	BrG	& 5.35 & 228.90 & 3.117 & 	2009-10-28	&	8.7	\\
HD 76943	&	BrG	& 0.50 & 233.34 & 1.161 & 	2012-03-10	&	9.34	\\
HD 78154	&	BrG	& 4.02 & 348.12 & 2.337 & 	2012-03-10	&	9.63	\\
HD 82328	&	BrG	& 2.59 & 145.18 & 5.777 & 	2012-03-10	&	10.37	\\
HD 82885	&	BrG	& 6.65 & 61.00 & 3.753 & 	2012-03-10	&	9.03	\\
HD 98231	&	BrG	& 1.50 & 196.14 & 0.271 & 	2012-03-10	&	9.48	\\
HD 100180	&	Ks	& 14.66 & 329.39 & 1.439 & 	2012-03-10	&	10.32	\\
HD 101177	&	Ks	& 8.79 & 248.08 & 0.879 & 	2012-03-10	&	10.12	\\	
\hline
\end{tabular}\\
{A sample of measurements of companions detected in the IRCAL FOV. The depth is the 5-$\sigma$ limit reached at 4--5$''$ range, where the contrast is highest. The full table is available online. \\
} \label{binarytab1}
\end{table}

\begin{table}
 \centering
\caption{Single Stars Observed with IRCAL}
\begin{tabular}{lccc}
\hline
Name & Filter & UT Date & Depth \\
 &  &  & (mag) \\
\hline
\hline
HD 166	&	BrG	&	2009-10-28	&	9.12	\\
HD 693	&	Ks	&	2010-08-03	&	8.29	\\
HD 1326	&	Ks	&	2009-10-28	&	8.75	\\
HD 1835	&	Ks	&	2010-08-05	&	9.07	\\
HD 3651	&	BrG	&	2009-10-28	&	7.96	\\
HD 4628	&	Ks	&	2010-08-03	&	8.69	\\
HD 4676	&	BrG	&	2009-10-28	&	7.9	\\
HD 4813	&	BrG	&	2010-08-05	&	8.95	\\
HD 5448	&	BrG	&	2009-10-28	&	8.89	\\
HD 7439	&	BrG	&	2010-08-05	&	9.28	\\
HD 10307	&	BrG	&	2009-10-28	&	8.23	\\
HD 10476	&	BrG	&	2010-08-05	&	8.99	\\
HD 11171	&	BrG	&	2010-08-05	&	8.21	\\
HD 11636	&	BrG	&	2010-08-05	&	9.02	\\
HD 13161	&	BrG	&	2009-10-28	&	9.2	\\
HD 13974	&	BrG	&	2009-10-28	&	8.69	\\
HD 14055	&	BrG	&	2009-10-28	&	8.72	\\
HD 16673	&	BrG	&	2010-08-05	&	9.04	\\
HD 17093	&	Ks	&	2009-10-28	&	8.61	\\
\hline
\end{tabular}\\
{Lick IRCAL observations with no companions detected. The depth is the 5-$\sigma$ limit reached at 4--5$''$ range, where the contrast is highest. The full table is available online.} \label{singletab1}
\end{table}

\subsection{Newly Identified Companions}

Most targets observed either had no detected companion or were binaries previously known in the literature. 
However, we have identified 5 new companions. We discuss each new system below.

{\it HD~1404:} This is an A2 star located 41~pc from Earth.
We detect a companion 6.7 magnitudes fainter at Ks and Br-$\gamma$ that over a three year period has remained at a projected separation of 6.6\arcsec\ with position angle (PA) of 144 degrees. The companion has Ks$\sim$11.2, or $M_{Ks}\sim$8.1. At an age of 450~Myr \citep{Vican:2012}, the \citet{Baraffe:1998} models suggest a mass of $\sim$0.17\msun and effective temperature of $\sim$3200~K (spectral type M4) for this companion. This A+M binary is separated by 271~AU and hosts a debris disk of radius 22~AU around the primary A star \citep{Thureau:2014}.

{\it HD~168151:} This system is an F star situated 32~pc from Earth. 
We detect a companion 3.3\arcsec\ from the primary in observations carried out in 2010 and 2012. The companion is 6 magnitudes fainter than the primary suggesting an absolute $M_{Br\gamma}$ magnitude of 7.4. The system's age is estimated to be 1.3~Gyr \citep{Vican:2012}. At that age, the \citet{Baraffe:1998} models predict a T$\sim$3400~K and 0.3\msun ($\sim$M3) dwarf would have that absolute magnitude.

{\it HD~140538:} This G-star had a previously known companion at a projected separation of 4.2\arcsec\ \citep{Eggleton:2008}. With our Lick survey, we have resolved the companion into two equal brightness components separated by 0.22\arcsec. At a distance of 15~pc, this separation corresponds to just over 3~AU. We estimate J, FeII, and Br-$\gamma$ magnitudes of 8.5, 7.9, and 8.1 for the pair, with one component being $\sim$0.1 mag fainter than the other. The system has various age estimates, but as suggested by \citet{Vican:2012}, we adopt the chromospheric age estimate of 3.6~Gyr. The \citet{Baraffe:1998} models suggest a mass of 0.25--0.3\msun ($\sim$M3) for each resolved component.

{\it HIP~42220:} This is a late-K star 17~pc away that has been previously suggested to be an astrometric binary \citep{Makarov:2005}. We detect a companion 0.4\arcsec\ from the primary with $\Delta$mag of 2.3--2.4 across J, H, and Ks. The JHKs magnitudes of the companion are 8.7, 8.2, and 7.9, respectively. The J--Ks color (0.8) is suggestive of an early to mid M-dwarf. No age estimate is available for HIP~42220 \citep{Vican:2012}, but K-stars in our sample have median ages of 2~Gyr and dispersion of 2.9~Gyr. From the absolute Ks magnitude (6.7) and an age range of 100~Myr--5~Gyr, we estimate the mass of the companion lies roughly in the range 0.3--0.35\msun.
The information listed in \citet{Makarov:2005} can be used with their Equation~2 to derive a limiting mass ratio for the unseen astrometric companion. In this case, assuming a circular orbit, total system mass of 0.9--1\msun, and minimum projection effects, we find our detected companion is massive enough to be consistent with the astrometric signature observed by \citet{Makarov:2005}.

{\it HD~110833:} This system is a K3+K3 binary 15~pc away. \citet{Raghavan:2010} list this as an equal mass spectroscopic binary with a 271-day period. 
\citet{Vican:2012} estimates an age of 2.2~Gyr for this system.
Our AO work reveals another star 1.6\arcsec\ from the primary with $\Delta$mag of 4.2--4.4 across J, H, and Ks. 
This projected separation amounts to $\sim$24~AU.
The absolute JHKs magnitudes of the companion are approximately 8.7, 8.4, 8.1, and the \citet{Baraffe:1998} models suggest effective temperatures of $\sim$3200~K and mass $\sim$0.17\msun for that age. The detected companion thus constitutes a mid-M dwarf in the system.

\section{Debris Disks and Multiplicity}

\begin{figure}
\begin{center}
\includegraphics[width=88mm,angle=0]{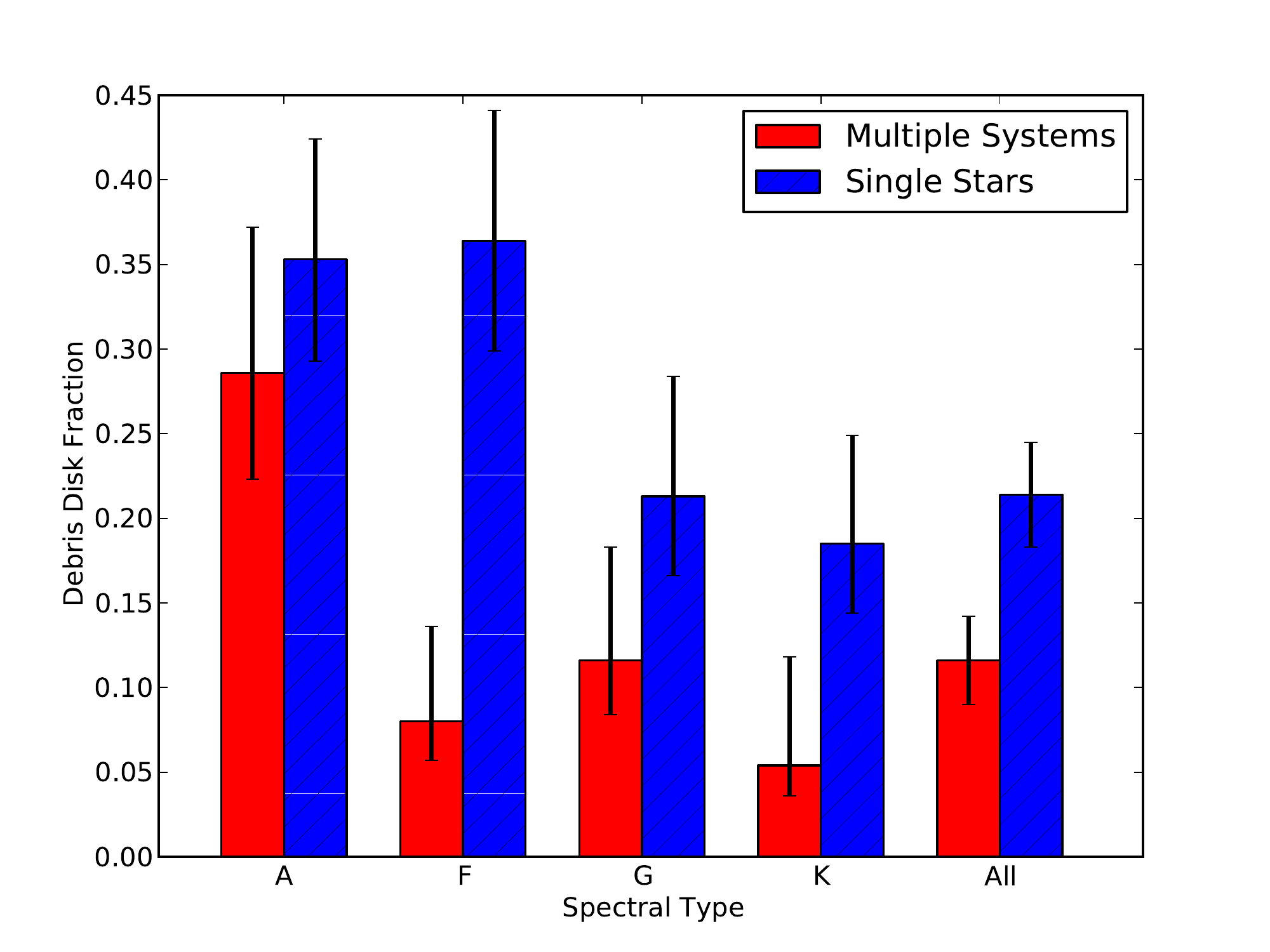}
\end{center}
\caption{Disk detection frequency among the DEBRIS sample, divided by spectral type and binarity.}
\label{fig:diskfrac}
\end{figure}

Systems in the DEBRIS sample have been observed with the Herschel spacecraft with PACS (100 and 160$\mu$m; \citealt{Poglitsch:2010}) and in some cases with SPIRE (250, 350, 500$\mu$m; \citealt{Griffin:2010}) and/or additionally at 70/160$\mu$m with PACS. Furthermore, most also have observations with IRAS, Spitzer-MIPS, AKARI, and WISE. Hence, the spectral energy distribution (SED) of these objects can be sampled in a broad wavelength range and any IR excess emission can be characterized and studied in detail.
Full descriptions on the observation strategy, source extraction, SED modeling, and selected results can be found in \citet{Matthews:2010} and other DEBRIS publications (e.g., \citealt{Lestrade:2012}, \citealt{Booth:2013}, \citealt{Thureau:2014}, Matthews et al., in prep; and references therein).
To summarize, stellar atmosphere models are first fit to data shortward of about 10$\mu$m to determine the photospheric flux. These predictions are then compared to the mid to far-IR photometry to look for the excesses indicative of emission from cool dust. Disk systems are those with at least one photometric 3$\sigma$ excess. To characterise the disk properties, a blackbody is fit to the star-subtracted SED and modified by a factor $(\lambda_0/\lambda)^\beta$ for wavelengths longer than $\lambda_0$ to account for inefficient grain emission at long wavelengths. 
For most cases, $\lambda_0$ is left as a free parameter, though in cases with limited photometry it is fixed (see details in Matthews et al, in prep).
We note, however, that a number of systems have poorly constrained temperatures or are required to have cold disks ($T_{dust}\leq$20~K), which could instead be a result of contamination from extragalactic sources \citep{Sibthorpe:2013,Krivov:2013,Gaspar:2014}.
For purposes of this paper, we include these $\sim$8-10 systems as part of our statistical analysis.
Among the DEBRIS sample, we find 76 systems with IR excess emission suggestive of circumstellar (or circumbinary) disks. 
In two cases (Fomalhaut and HD~223352), two stars in the same system each show IR excess which implies 78 disks among individual components the sample (see Section~\ref{binarydisk}). 

\subsection{Global Statistics}

Among the DEBRIS sample, there are 21 multiple systems which host circumstellar disks. There are several ways to approach the statistics of stellar multiplicity and disk incidence in our sample. 
We first consider the disk detection frequency among both single and multiple stars by spectral type.
We note that our ability to detect a disk, or, more accurately, detect an infrared excess, will depend on the spectral type because our survey is flux-limited. 
The distribution of distances for single and multiple star systems is similar so no distance biases are introduced.
We note that multiple systems, regardless of the number of components, are counted only once. 
Considering the entirety of our 449-star sample, the 76 disk systems identified suggest an overall detection frequency of $17\pm2$\%. 
For single stars, this is $55/261=21\pm3$\% and for multiples, $21/188=11\pm3$\%.
Figure~\ref{fig:diskfrac} summarizes the disk detection frequency in the DEBRIS sample as a function of spectral type and binarity. 
The detection of disks around A-stars is comparable regardless of multiplicity (35$^{+7}_{-6}$\% for singles and 29$^{+9}_{-6}$\% for multiples; see also \citealt{Thureau:2014}). 
Among the lower mass stars, multiples host fewer disks than single star systems. 
For the FGK sample, we find single stars have a disk detection frequency of 25$^{+4}_{-3}$\% while multiple systems have 8$^{+3}_{-2}$\%.
A two-sided Fisher test for the A-stars reveals a p-value of 0.64, which implies any different in detected disk frequency is not statistically significant.
In contrast, for the FGK sample the Fisher test returns a p-value of $3.5\times10^{-4}$ implying a statistically significant difference in the detected disk frequency between single and multiple star systems.

\begin{figure}
\begin{center}
\includegraphics[width=88mm,angle=0]{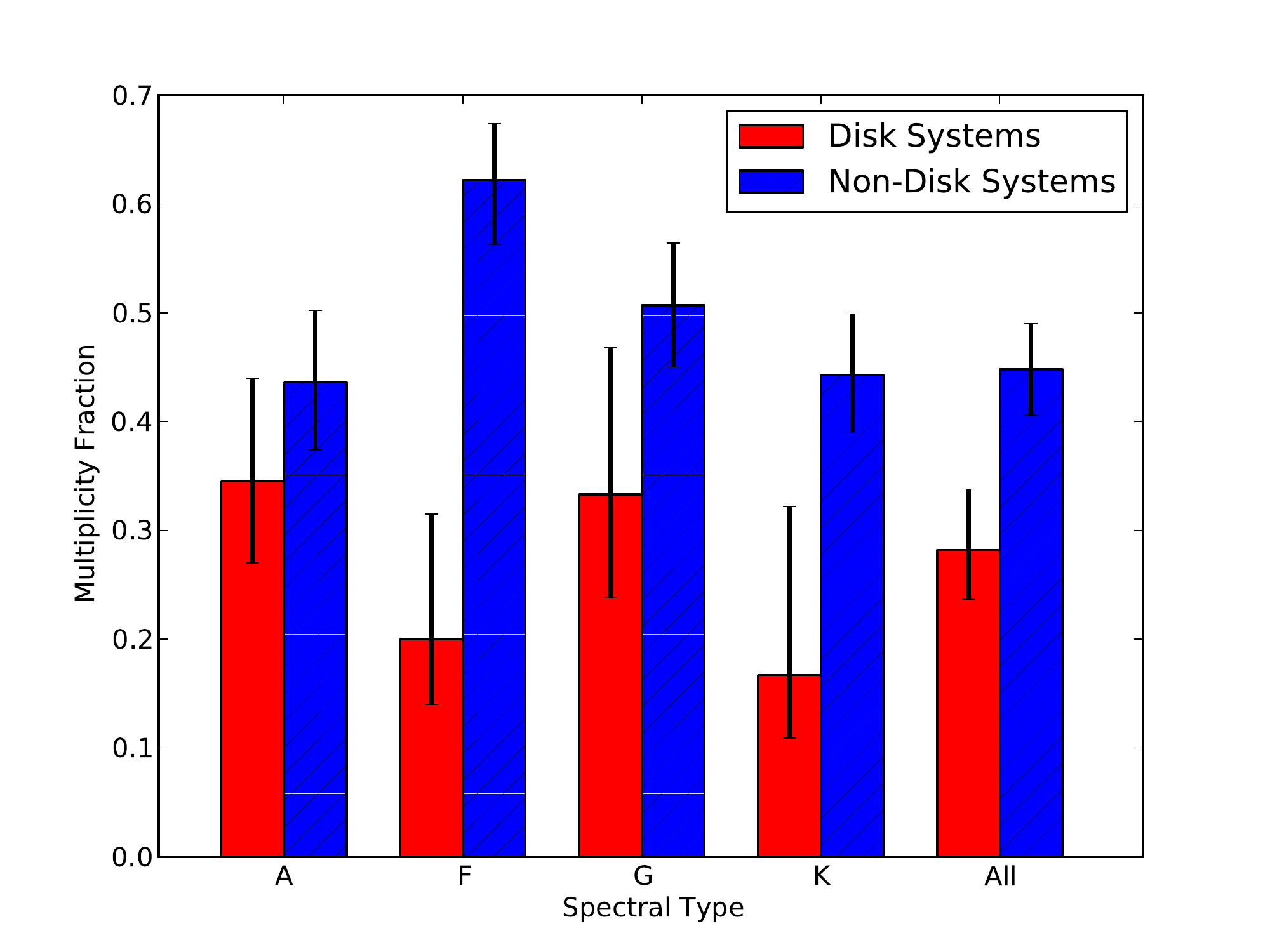}
\end{center}
\caption{Multiplicity fraction among the DEBRIS sample, divided by spectral type and presence or absence of detected disks.}
\label{fig:binfrac}
\end{figure}

An equivalent way of examining the frequency of disks around multiple-star systems is to consider the multiplicity fraction around dusty stars and those not bearing disks. 
This fraction is the ratio of the number of binary or multiple systems compared to the total number of systems in the disk or diskless samples; that is, the individual components within a multiple system are not considered.
Figure~\ref{fig:binfrac} illustrates the multiplicity fraction as a function of spectral type. 
Again, A-stars show similar multiplicity fractions regardless of whether we consider the disk and non-disk sample. Disk-bearing stars among the lower mass stars, however, have lower multiplicities. That is, it is rarer to find an FGK disk-bearing system that is also a binary or multiple.

In addition to the incidence of infrared excesses among the DEBRIS sample, we examine properties of the disk, namely the distribution of dust temperature and the disk luminosity divided by the stellar luminosity ($L_{IR}/L_{bol}$; or fractional luminosity), to see if any differences or trends are evident in the sample.
A KS test reveals no difference between single and multiple stars in terms of dust temperature (p=0.9) or $L_{IR}/L_{bol}$ (p=0.5), as was also demonstrated for A stars by \citet{Thureau:2014}. This holds regardless of whether we consider the full sample or break it as A or FGK stars: for dust temperatures we find p=0.3 for A singles vs multiples and similarly p=0.9 for FGK systems. For $L_{IR}/L_{bol}$, we find p=0.3 and 0.4 for A and FGK systems, respectively.
For systems with detected disks, the basic properties of these disks don't appear to be correlated with the multiplicity of the system. 

\subsection{Disk Frequency Around A and FGK Stars}

\begin{figure}
\begin{center}
\includegraphics[width=88mm,angle=0]{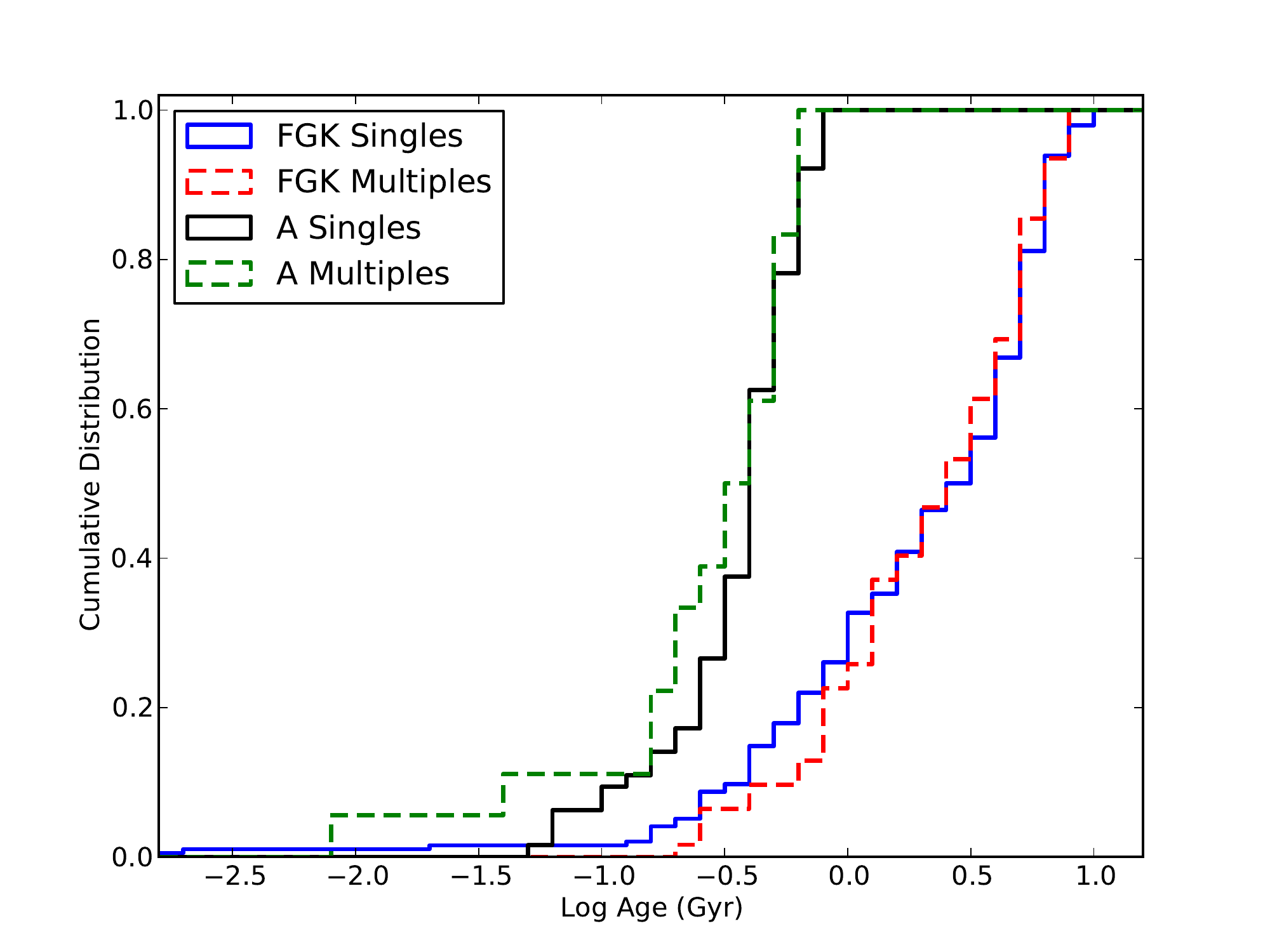}
\end{center}
\caption{
Cumulative age distribution for A and FGK stars in both single and multiple star systems.
}
\label{fig:agedistribution}
\end{figure}

The frequency of disks around A stars, as well as their properties, is the same regardless of multiplicity. The situation is different for FGK stars in that while the disk properties are similar, the frequency of disks is lower among the multiple systems. 

The dynamical effects of a second star can serve to disrupt the disk, accelerating its dispersal. As such, the disk detection frequency among multiples should be lower, or equivalently, disk-bearing systems should more likely be single stars.
The discrepancy between the A-star and FGK sample, however, could be a result of a selection bias due to their ages.
Examining ages from \citet{Vican:2012}, we find dusty A-stars in the DEBRIS sample have a median age of about $\sim$0.3~Gyr, whereas the dusty FGK-stars are older, at a median age of $\sim$4~Gyr.
On the other hand, the binary fraction of disk-bearing systems for $<$1~Gyr and $>$1~Gyr FGK stars is $29^{+14}_{-9}$\% and $21^{+9}_{-6}$\%, respectively. While we expect older FGK stars to be less likely to host disks, the multiplicity difference we observe is not statistically significant. 

We also considered the age distribution of A, F, G, and K stars for both single and multiple stars regardless of the presence of disks. If the ages were different as a function of multiplicity, these could account for the observed difference in the disk frequency. However, both single and multiple star distributions look very similar (see Figure~\ref{fig:agedistribution}). The KS test does not show evidence that singles and binaries of any spectral type are drawn from separate age distributions.

As a further test to ascertain the difference between A and FGK binaries with disks, we constructed samples of systems eliminating either very wide binaries ($>$500~AU) or very close binaries ($<$1~AU). 
In both cases, these companions tend to be more common for A stars than lower-mass stars and could contribute to the discrepancy (see Figure~2 in \citealt{Duchene:2013}).
These systems have little effect on the disks Herschel is sensitive to. 
However, in our sample there are only a handful of these wide or close systems and as such the statistics do not appreciably change when eliminating these. 
In fact, the distribution of periods or semi-major axes between A and FGK multiples among this sample is not appreciably different.

The discrepancy between A and FGK disk-bearing binaries does not appear to be a result of binary properties or an age bias in our sample.
A possible explanation is that we are missing a significant number of companions among A-stars, as suggested by the higher multiplicity fraction ($>$50\%) in \citet{Duchene:2013}. However, the A-stars in the DEBRIS sample are generally well-studied and we would require 6--7 additional binaries, all without detected disks, to yield a comparable difference between disk-bearing and non-disk bearing systems as observed among the FGK stars.

\subsection{Disks in Binary and Multiple Systems}

\begin{figure}
\begin{center}
\includegraphics[width=88mm,angle=0]{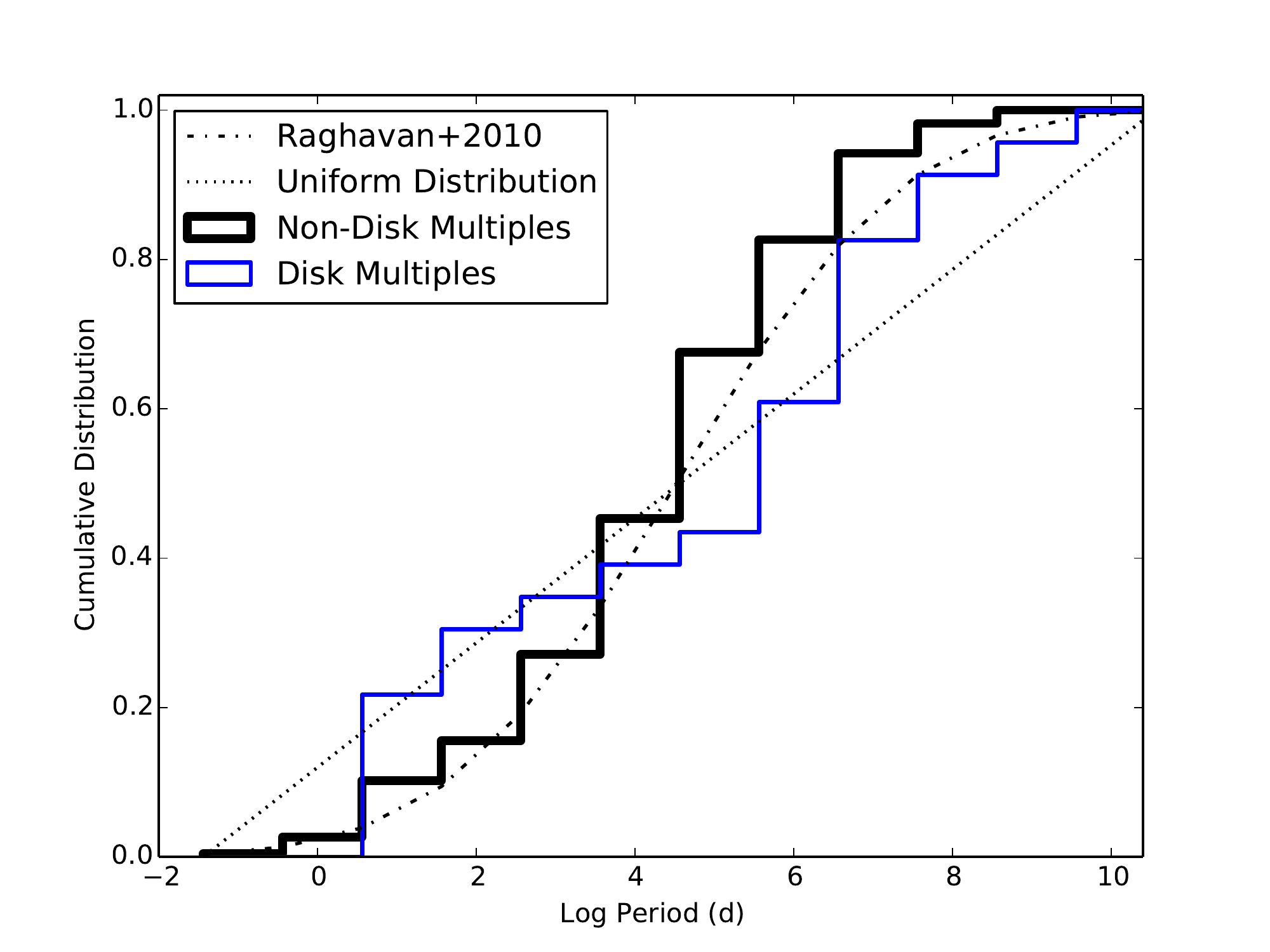}
\end{center}
\caption{Cumulative distribution of calculated and measured orbital periods for multiples. Black shows the full sample, while blue are disk-bearing binaries. The \citet{Raghavan:2010} distribution is shown as a dashed line and a flat distribution is shown with the dotted line.}
\label{fig:cdf}
\end{figure}

As previously mentioned, the binary period distribution of the full DEBRIS sample is consistent with prior studies of stellar multiplicity \citep{Eggleton:2008,Duquennoy:1991,Raghavan:2010}. 
Figure~\ref{fig:period} highlights the period distribution of disk-bearing systems, which appears flat in comparison to the full sample. 
Figure~\ref{fig:cdf} shows the cumulative distribution of periods for the non-disk and disk binaries compared to the distribution of \citet{Raghavan:2010} and a flat distribution across all periods. 
At a glance, among the relatively low number of disk-bearing binaries there appear to be fewer of these with periods between $10^3$ and $10^5$ days.
However, the period distribution of disk-bearing DEBRIS systems does not differ significantly from that of the non-disk sample. A KS test returns a p-value of 0.09 when comparing these two distributions. 
For us to assert that the two samples to are drawn from different distributions we require a p-value of 0.05 or lower.
Hence, we cannot claim that the distribution of disk and non-disk binaries differ from each other or from the general distribution of \citet{Raghavan:2010} or from a flat, uniform distribution. 
While the various distributions look tantalizingly different at first glance, we cannot rule out they are drawn from the same underlying distribution with the limited statistics offered by this survey.

\begin{figure}
\begin{center}
\includegraphics[width=88mm,angle=0]{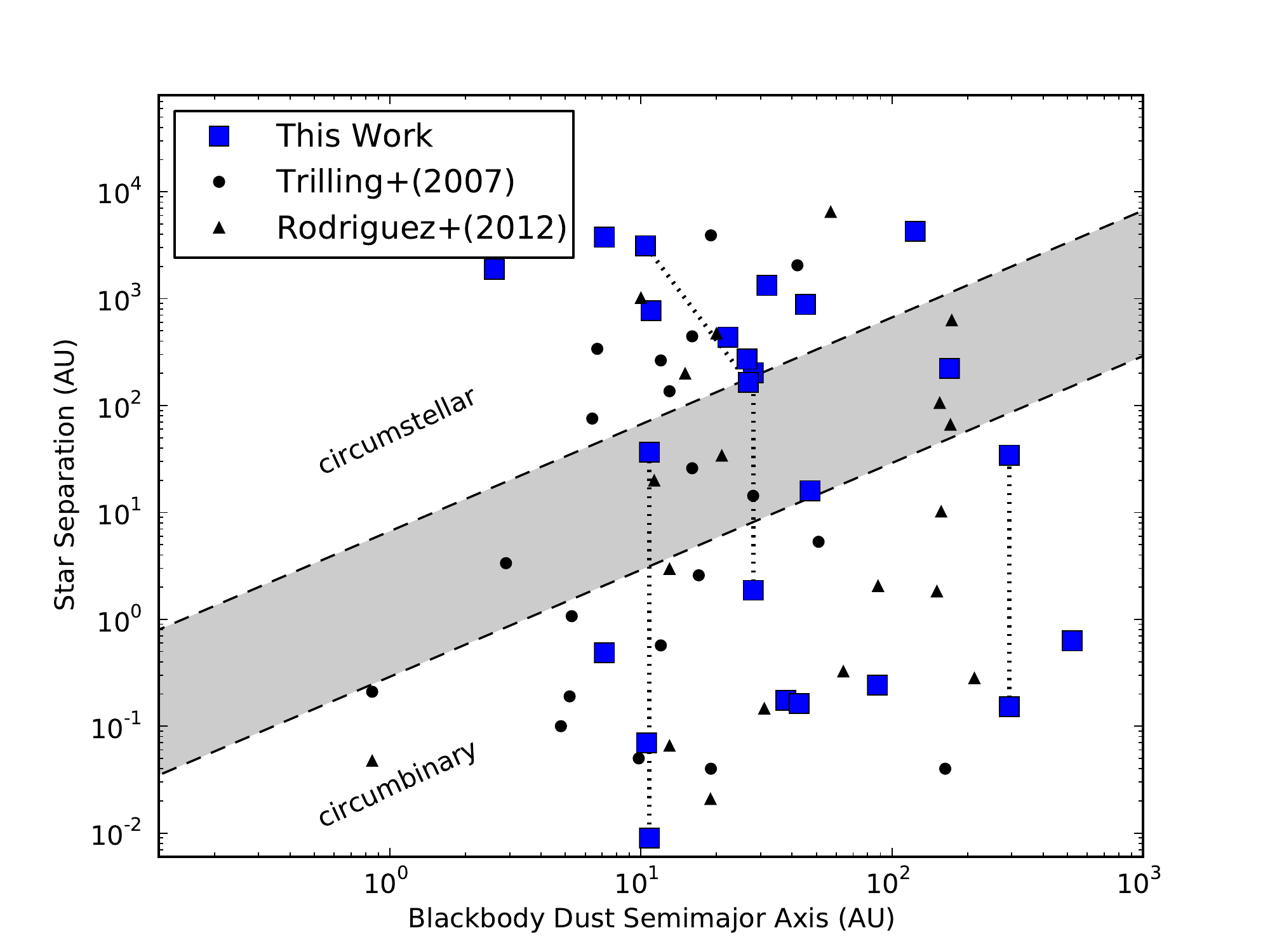}
\end{center}
\caption{Stellar separation vs blackbody dust semimajor axes for the dusty multiple systems in the DEBRIS sample and from other surveys in the literature. The grey denotes the approximate region at which gravitational effects can perturb the disk. Triples where the dust lies between the AB and C pair are connected with a dotted line. HD~223352 is the system with the slanted line as the system hosts two separate debris disks. 
With stellar separations exceeding 150~kAU, the three components of HD~216956 (Fomalhaut) are beyond the plot range in this figure. }
\label{fig:stardustsep}
\end{figure}

Another way to examine the disk-bearing binaries is by examining the stellar separation.
Figure~\ref{fig:stardustsep} shows the stellar separation compared to the location of the dust in the system for multiples in this work, as well as others from \citet{Rodriguez:2012} and \citet{Trilling:2007}.
For the dust, we plot the semimajor axis of the dust assuming it is composed of large, blackbody grains. Some disks are spatially resolved and discussed elsewhere (e.g., \citealt{Kennedy:2012a, Kennedy:2012b,Booth:2013}).
For A-stars, the disk location for resolved systems tends to be a factor of $\sim$1--2.5 times that estimated by the assumption of blackbody grains; FGK disks may be larger (see \cite{Pawellek:2014}). As such, some of the systems may be shifted rightward in the plot. 
For clarity and consistency, we use only the dust semimajor axis as derived from the SED rather than any resolved radius. 
Some systems may be shifted upward because of projection effects (or downward if observed near periastron) in the plot, but this tends to be a small correction (e.g., \citealt{Dupuy:2011}).
Some multiples are connected by dotted lines corresponding to the stellar separations. For example, a triple may consist of a close binary surrounded by a dust disk and a more distant stellar companion. Both the separation of the close binary and the more distant companion would be plotted and connected in Figure~\ref{fig:stardustsep}.

Figure~\ref{fig:stardustsep} also shows a grey region representing the area where the gravitational influence of a companion would disrupt the disk. This is done by computing the critical semimajor axis using the relationships in \citet{Holman:1999}. This is the distance at which a test particle would survive for less than $10^4$ times the binary orbital period. 
For these relationships, we adopt 0.5 for the mass ratio and 0.4 for the eccentricity and derive critical semimajor axes values of 0.15 and 3.4 times the stellar semimajor axis (see Tables~3 and 7 in \citealt{Holman:1999} for values for other mass ratios and eccentricities). 
Systems located in this region would have their disks quickly cleared out by the stellar companion. 
Of the 26 DEBRIS components plotted, 4 (HD~223352, 99~Her, HIP~14954, HIP~73695) lie in this unstable region, or 15$^{+10}_{-5}$\%.
If we generate a random sample populating the diagram, either with a uniform distribution or a distribution following the period distribution of \citet{Duquennoy:1991} and with a disk population uniformly spread between disk radii of 0.12 to 1000 AU, we find that $\sim$20\% of components lie in the unstable area.
Errors due to unknown inclinations or dust grains deviating from blackbodies were not included in this simulation, but their effects are expected to be minor (see above and \citealt{Dupuy:2011}).
This suggests the location of the dust in the system may not be strongly correlated with the location of the stellar companion.

\begin{figure}
\begin{center}
\includegraphics[width=88mm,angle=0]{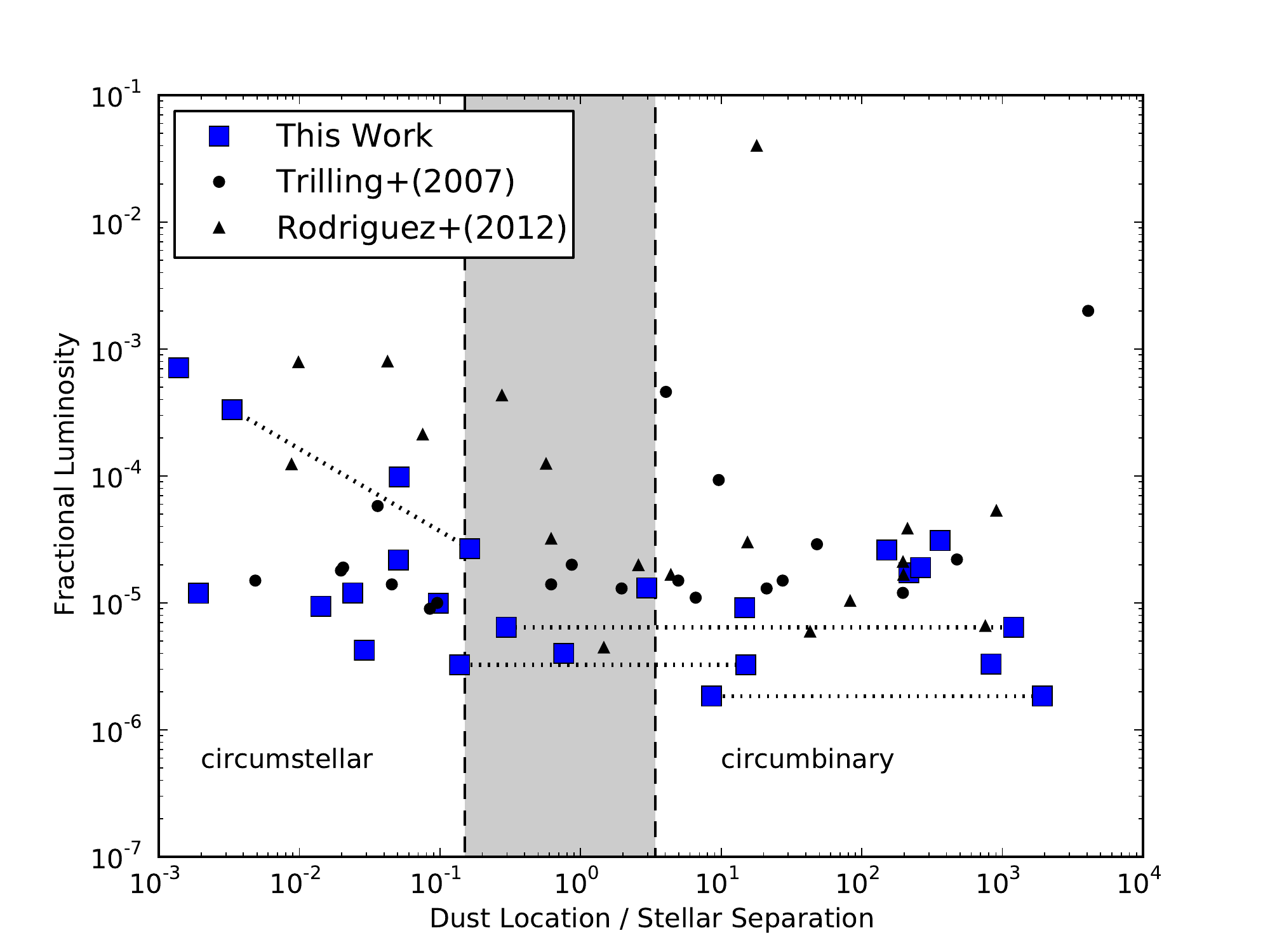}
\end{center}
\caption{Ratio of dust to stellar separation compared with $L_{IR}/L_*$ for dusty multiple systems. The grey denotes the approximate region at which gravitational effects can perturb the disk. As before, triples are connected and HD~223352 is the system connected with a slanted line as two separate debris disks are present in the system. HD~216956 (Fomalhaut) has a ratio $<10^{-3}$ and is not shown in this figure.
The system with fractional luminosity $\sim$4\% is BD+20~307.}
\label{fig:tausep}
\end{figure}

As previously mentioned, the fractional luminosity, $L_{IR}/L_{bol}$, of multiples with detected disks is not significantly different from that of single stars within the DEBRIS sample. Figure~\ref{fig:tausep} compares this fractional luminosity against the ratio of the dust semimajor axis and the stellar separation. Systems located towards the left of the plot are circumstellar in nature where the dust orbits a single star, yet there is a distant stellar companion in the system. Systems towards the right are circumbinary in nature in that the dust surrounds a pair of stars. Again, the unstable zone is highlighted as discussed above.
The spread in fractional luminosity is comparable between the circumstellar disks (dust location / stellar separation ratio $<0.1$) and the circumbinary disks (ratio $>3$).
Combined with the results of Figures~\ref{fig:cdf}~and~\ref{fig:stardustsep}, this suggests the properties of disks around disk-bearing binary stars do not, in general, depend strongly on the orbital properties of the binary.

\subsection{Individual Binary Debris-Disk Systems}\label{binarydisk}

In this section, we highlight a handful of disks around binary or multiple star systems. These stand out in our sample as we describe below.

{\it HD~223352:} As detailed in \citet{Phillips:2011}, HD~223352 is a quadruple system in which two components, A and C (HD~223340), host circumstellar disks. 
The AB pair is separated by 3.9\arcsec~(164~AU) and the C component is much farther away at 74\arcsec~(3100~AU). Recent observations have resolved the B component as two separate stars \citep{deRosa:2011}. The two disks in the system are located around the A and C stars, with estimated blackbody dust locations of 27 and 10~AU. 
The ratio of the dust to stellar separation for the AB pair is $27/164=0.16$ placing it just inside the unstable region in Figures~\ref{fig:stardustsep} and~\ref{fig:tausep}, whose upper bound is set at a ratio of 0.15.
For Figures~\ref{fig:stardustsep} and~\ref{fig:tausep}, we have taken care to connect the two disk-bearing components together as each disk is associated with a different star separation within the system.

Very few debris disk systems are known in which two stars host individual disks. Disregarding $\sim$10--20~Myr-old (and younger) systems whose disks may be primordial in nature (see, e.g., \citealt{Rodriguez:2014}, and references therein), the only other multiple system known to host two disks is that of Fomalhaut, also part of this survey (see \citealt{Kennedy:2014}). 
As a member of AB~Dor \citep{Malo:2013,Zuckerman:2011}, HD~223352 has an age of $\sim$100~Myr and  represents a rare opportunity to study the development and evolution of disks around a pair of co-eval stars.

{\it HD~165908:} Commonly known as 99~Her, this system has been described in detail in \citet{Kennedy:2012a}. With a stellar separation of $\sim$16~AU and a blackbody disk dust location of $\sim$47~AU, it is one of the unstable disk systems in our study. However, the disk has been resolved by Herschel to be located $\sim$120~AU from the binary pair, which places the system beyond the grey unstable zone in Figure~\ref{fig:stardustsep}. A more detailed discussion of the disk and its interaction with the binary can be found in \citet{Kennedy:2012a}.

{\it HIP~14954:} An F8 primary orbited by a widely separated M star at $\sim$223~AU. Like 99~Her above, HIP~14954 appears to have dust located in an unstable configuration given the blackbody dust location of $\sim$164~AU is comparable to the stellar separation. 
\citet{Roberts:2011} provides orbital parameters for this system: a period of 2029 years, a semi-major axis of 9.87$''$ (223~AU), and eccentricity of 0.26.
Using these orbital elements, the relationships of \citet{Holman:1999} predict critical semimajor axes of 59 and 708~AU. That is, orbits whose semimajor axis is smaller than 59~AU (or larger than 708~AU) are stable on timescales longer than $10^4$ times the stellar period ($\sim$20~million years in this case). The fact that the dust is located beyond 59~AU suggests it should have been quickly disrupted or instead has been recently produced. 
However, this system shows possible confusion in the Herschel data and the infrared excess could be originating from a background source.

{\it HIP~73695:} A triple system in which the primary is orbited by a 0.3-day binary. The binary orbits the primary with a period of 206 years and eccentricity 0.55 \citep{Eggleton:2008}; this amounts to a semi-major axis of about 37~AU. 
The dust is around the primary at a blackbody dust location of $\sim$11~AU. 
The location of the dust is uncertain as the infrared excess is detected only at 70$\mu$m.
With the binary properties listed in \citet{Eggleton:2008}, we estimate critical semimajor axes of 4 and 138~AU. As with HIP~14954, the dust appears to be located beyond the inner critical semimajor axis. 
Additional information would be needed to make a definitive statement on the stability of the disk in the system.

\section{Conclusions} \label{conclusions}

The DEBRIS sample consists of 449~AFGKM systems observed with Herschel to search for circumstellar disks. In this study, we have examined these stars for signatures of stellar multiplicity. 
In this sample, 42\% of the targets are binary and higher order stellar multiples, as noted by others in the literature and our own adaptive optics observations (Table~\ref{tab:bin1}). This sample allows us to examine the influence of stellar multiplicity on the incidence and properties of debris disks.

Disk systems among the sample are summarized in Matthews et al.\ (in prep), \citet{Thureau:2014}, and Sibthorpe et al.\ (in prep).
Despite the large number of systems considered, the number of disk-bearing systems (78 across all spectral types) is lower than considered in \citet{Rodriguez:2012}. This is a result of the way these two samples were constructed. Nevertheless, an advantage to this study is that it allows us to consider both the disk frequency (as in \citealt{Trilling:2007}) and the binary/multiple frequency (as in \citealt{Rodriguez:2012}) simultaneously for the same sample. 
Our results are, at a glance, similar to both prior studies.
However, an examination of the period distribution of the multiples in the sample shows no statistically significant difference with regards to whether or not these systems host debris disks and instead follows more closely the distribution previously reported for stellar multiples in general (e.g., \citealt{Duquennoy:1991, Raghavan:2010}).

Unlike prior studies, the larger sample allows us to break up the results as a function of spectral type. 
Among A-type stars, as primarily examined in \citet{Trilling:2007}, \citet{Rodriguez:2012}, and \citet{Thureau:2014}, disks are detected just as frequently among single stars as in multiple star systems.
Equivalently, disk and non-disk A-star systems show comparable multiplicity. The results for FGK stars, however, differ substantially from the A stars. In FGK stars, the presence of a second star reduces the likelihood to find a circumstellar or circumbinary disk in the system. 
We have ruled out potential biases due to age or binary properties, but have been unable to find a convincing mechanism for this difference compared to A stars. 
Having a significant number of missing binaries among A-stars remains a plausible explanation.
A larger sample of well-characterized FGK stars may be needed to further examine the disk detection frequency as a function of stellar multiplicity.

Among the disk-bearing multiples, both circumstellar and circumbinary disk systems exist. 
The basic properties of detected disks in binary systems, namely the temperature of the dust and the fractional luminosity, are comparable to those in single star systems. 
There are a few systems (described in more detail in Section~\ref{binarydisk}) in which the stellar separation and the location of the dust (assuming large blackbody grains) are comparable to each other. This would result in an unstable scenario in which the gravity of the stellar companion would readily disrupt the disk. However, limitations in our knowledge of the orbital parameters, namely the projection on the sky, as well as the possibility of inefficient grain emission in the disk, can conspire to make a system appear to have an unstable configuration when it does not.

We interpret these results as follows.
Any sample of binary stars dominated by FGK stars (as is this one) will in general demonstrate the period distribution of stellar multiples of \citet{Duquennoy:1991} and \citet{Raghavan:2010}. 
The distribution of such a sample will not vary significantly between systems with detected dust disks and the general population.
A binary star hosting a circumstellar or circumbinary disk will exhibit properties consistent with other (non-disk bearing) binaries and the dust properties in the system will be comparable to those of single stars with disks. However, the gravitational influence of the companion star(s) will accelerate the collisional evolution of the disk. As such, a sample of binary stars will have a detectable excess for a shorter period of time.
Hence, disks will be less readily detected among a sample of old binary or multiple stars than single stars of comparable spectral types and ages. However, given the stochastic nature of debris disk formation as a function of time, this signature is diminished among samples dominated by stars with ages larger than a few hundred million years.

The increasing number of planets found orbiting pairs of stars shows that planet formation in binary systems is not an uncommon product of planetary evolution. Binary and multiple star systems are plentiful and our work suggests that the processes behind disk and planet formation are similar between single and multiple star systems.

\section*{Acknowledgments}

We thank H. M. Butner's students, D. Simonson, and D. Trelawny for their assistance at Lick Observatory and Mark Booth for comments on a draft of this article.
We thank the anonymous referee for a prompt review and constructive suggestions.
This research has made use of the Washington Double Star Catalog maintained at the U.S. Naval Observatory, and the SIMBAD database and the VizieR catalog access tool, operated at CDS, Strasbourg, France.
D.R.R. acknowledges support from project BASAL PFB-06 of CONICYT, a Joint Committee ESO-Government of Chile grant, and FONDECYT grant \#3130520.
This work was supported in part by NASA through a contract (No.\ 1353184, PI: H. M. Butner) issued by the Jet Propulsion Laboratory, California Institute of Technology under contract with NASA.
GMK acknowledges support from the European Union through ERC grant number 279973.

\label{lastpage}

\end{document}